\documentclass{aa}             % <-- "class" anstelle von "style"
\usepackage{amssymb,graphicx}  % <-- "graphicx" fuer Bilder, "amssymb"
\begin{document}

\title{The formation of the coronal flow/ADAF}
\author{ E. Meyer-Hofmeister and F. Meyer}
\offprints{Emmi Meyer-Hofmeister}
\institute{Max-Planck-Institut f\"ur Astrophysik, Karl-
Schwarzschildstr.~1, D-85740 Garching, Germany
} 

\date{Received: / Accepted:}

\abstract{
We develop a new method to describe the accretion flow in the corona
above a thin disk around a black hole in vertical and radial extent.
The model is based on the same physics as the earlier one-zone
model, but now modified including inflow and outflow of mass,
energy and angular momentum from and towards neighboring zones.
We determine the radially extended coronal flow for different mass
flow rates in the cool disk resulting in the truncation of the thin
disk at different distance from the black hole. Our computations show
how the accretion flow gradually changes to a pure vertically extended
coronal or advection-dominated accretion flow (ADAF). Different
regimes of solutions are discussed. For some cases wind loss causes 
an essential reduction of the mass flow.

}
\titlerunning {}
\maketitle
% 4 figures:H4183f1.eps,... H4183f4.eps
%_____________________________________

\section{Introduction}

The new fascinating results from XMM-Newton and the Chandra X-ray
Observatory allow a deeper insight into the physical processes in
many astrophysical objects, on scales ranging from binary stars
to galaxies. One of these topics is the
accretion onto black holes. This can happen in form of an
advection-dominated accretion flow (ADAF) or via a standard 
geometrically thin Shakura-Sunyaev accretion disk. Since an ADAF is
only possible in the inner region around the central accretor one has
in many objects an advection-dominated accretion flow in the inner
region, fed by the mass flow through a standard disk at larger
distances from the black hole (for a recent review on ADAF models see
Narayan 2002). The situation is the same for accretion onto black
holes in galactic X-ray binaries and onto supermassive black holes in
active galactic nuclei (AGN).

The spectra arising from
these two different ways of accretion are very different, a soft
multi-temperature spectrum from the black-body like accretion disk
and a hard power law spectrum from the Comptonizing much hotter gas
in an ADAF (one observes the combination of both originating at
different distances from the black hole). The observed change of the
spectral type, the soft/hard
spectral transitions in X-ray binaries, were successfully modeled
as related to a changing mass flow rate by Esin et al. (1997, 1998),
thereby supporting this picture of an ADAF in the inner region
surrounded by a standard disk. This means that at a certain distance
from the black hole the accretion mode changes from the flow via a
thin disk to a hot gas flow. The distance where this happens
depends on the mass flow rate from outside. 

Meyer, Liu and Meyer-Hofmeister (2000) developed a model for a corona
above a geometrically thin standard disk around a black hole (basically
the same physics as already discussed earlier for disks of dwarf nova
systems, where the compact
object is a white dwarf (Meyer \& Meyer-Hofmeister 1994)). The
corona is fed by gas which evaporates from the cool thin disk 
underneath. An equilibrium establishes between the cool
accretion stream and the hot flow. The efficiency of this process
increases towards the black hole. This means that at a certain
distance all matter is evaporated and the disk is truncated. From this
distance on inward all gas is in the hot flow and proceeds towards
the black hole as an advection-dominated flow.
This original model is a simplifying description of the 
two-dimensional accretion flow (of vertical and radial extent with a
free boundary condition on a radially extended surface).
The resulting evaporation efficiency as a function of distance
(and of the central black hole mass) allows to determine the location
at which the thin disk is truncated. 

For some X-ray binaries in quiescence the location of
the inner edge of the disk
can be deduced from the orbital velocity there, which is inferred from
the observed $\rm{H}_\alpha$ line profiles. This requires that the
disk temperature at the innere edge does not exceed the ionization
temperature of hydrogen. Using the ADAF
model for the fit to the spectra of soft X-ray transients as e.g. A620-00,
V404 Cyg and Nova Mus 1991 the mass accretion rates were derived (Narayan
et al. 1996, 1997). If one compares these results with the disk
truncation radius from the evaporation model the agreement is
reasonable (Liu et al. 1999, Meyer-Hofmeister \& Meyer 1999).
The change from disk accretion to an
ADAF was also investigated for several low luminosity AGN and
elliptical  galaxies (Quataert et al. 1999, Gammie et al. 1999,
Di Matteo et al. 1999, 2000), also the theoretically expected
disk truncation (Liu \& Meyer-Hofmeister 2001).
The results confirm that the truncation of the thin disk is located
at smaller distance from the black hole for higher mass flow rates
in the thin disk. Only for low luminosity AGN the observed spectra
seem to demand a disk truncation at radii too small for the accretion rate 
(Quataert et al. 1999). But this apparent discrepancy might be
resolved by the effect of
magnetic fields from a dynamo in the underlying disk on the coronal
gas flow (Meyer \& Meyer-Hofmeister 2002).
 
We now present work which is an essential step beyond the one-zone
model. We develop a new method to describe the accretion flow in the
corona above a thin disk in its vertical and radial extent. The
model is based on the same physics as in the one-zone model, but now modified
including inflow and outflow of mass, energy and angular
momentum from and towards neighboring zones. In the earlier model 
inflow from outward regions was neglected. But this is
necessary if one considers regions inside the evaporation maximum (in the
one-zone model at about 300 Schwarzschild radii, compare Fig. 3 in
Meyer et al. 2000). 
For the inner regions it is also important to take into account
different ion and electron temperature and, in the case of a high
mass flow rate in the thin disk, the effect of Compton
cooling of coronal electrons by disk photons.
Both were included in the one-zone model description by Liu et
al. (2002), where emphasis was put on evaluation of the
coronal flow in the case of high accretion rates as in narrow-line
Seyfert 1 galaxies. 

The co-existence of hot and cold gas around galactic black holes and
in AGN was also investigated by R\'o\.za\'nska and Czerny (2000a, b).
Different ion and electron temperatures were already included. In
their work the physical picture is basically the same as in ours,
but the results differ in detail. The investigation focuses on the
innermost region near the black hole. In earlier work
(Meyer-Hofmeister \& Meyer 2001) we discussed the difference of the
results. Processes of evaporation very close to the black hole were
studied by Spruit \& Deufel (2001).

In Sect. 2 we discuss the physics of interaction between the hot corona
and a cool disk underneath. We introduce the modifications necessary
for a consistent treatment of the mass, energy and angular momentum
flow. We describe the procedure how one finds the solution for the 
two-dimensional accretion flow, put together from the separately
computed structures for different radii, and
we present the new computational results (Sect.3). This
multi-zone model describes the increase of the mass flow in the hot
corona towards the black hole. A part of the matter is lost in a wind
from the hot corona. In Sect. 4 we discuss the different regimes
of solutions,  wind loss from the coronal flow and the consequences of 
a varying mass flow rate in the thin disk on spectral transitions.

\section{The physics of the interaction of the hot corona and a cool
disk underneath}
If we imagine hot gas above cool matter in a disk below both layers
interact. This interaction happens in the gravitational potential of
the central star (as earlier considered near the white dwarf in
binaries, a ``siphon flow'', Meyer \& Meyer-Hofmeister 1994).
The hot corona conducts heat downward by electron
conduction. At the bottom the temperature decreases from its high
coronal value to a low chromospheric value, heat conduction becomes
ineffective and the thermal heat flow has to be radiated away. The
efficiency of radiation (in the optical thin case) depends on the
square of the particle number density. If this density is too small
the material will heat up and increase the density in the corona. In
this way the corona of a given temperature will ``dig'' itself so deep
into the chromospheric layers that a density is reached which is able
to radiate away the downward heat conduction and an equilibrium 
establishes between the cool accretion stream and the hot
flow. The final density in the corona is determined by the pressure
equilibrium at the interface. The hot flow continuously drains mass
from the corona towards the central object. This is resupplied by
evaporation from the surface of the cool disk as the corona tries to
restore the density to the stationary level.

\subsection{The modeling of the corona above the thin disk in vertical
and radial extent}

To describe the structure of the corona above the cool disk we take
the standard equations of viscous hydrodynamics: conservation of
mass, the equations of motion and the first law of thermodynamics.
We want to determine the vertical structure of the corona at different
distances $r$ from the black hole with emphasis on the inner accretion
regions. 
For the earlier ``one-zone model'' only the zone at the edge of the thin disk
was considered. Since the evaporation efficiency increases steeply in
radial direction inward this approach gives already good results (but 
only for distances from the black hole where the evaporation
efficiency indeed increases inward). We now want to investigate the coronal
structure in its full extension in two dimensions. We determine the
vertically extended coronal structure for a series of
successive radial zones. In this multi-zone model the divergence
in radial direction is replaced by inflow/outflow of mass and angular
momentum in the zones (in the former one-zone model this could only
be taken into account approximately).

\subsection{The equations}

We use cylindrical coordinates $r,\varphi, z$ with
the $z$-axis perpendicular to the disk midplane.
We consider stationary and azimuthally symmetric flows.
We use basically the same equations as for the one-zone model, also
used in Liu et al. (2002), but modified to take
mass, energy and angular momentum inflow and outflow in its radial
dependence into account. 

In the following we list these five ordinary differential equations.
The dependent variables are the vertical mass flow $\dot m=\rho v_z$
($\rho$ density, $v_z$ vertical flow velocity), the vertical 
component of the heat flux $F_c$, the pressure $P$ and the ion and
electron temperature $T_i$ and $T_e$, the independent variables are
$r$ and $z$. 

We use the following equation of state  

\begin{equation}
P=n_{i}k T_{i}+ n_{e}k T_{e}\approx
\frac{\Re \rho}{2\mu_0} (T_{i}+T_{e}),
\end{equation}
with $n_{i}, n_{e},T_{i}, T_{e}$ ion and electron number density and
respectively temperature. $\rho_i=\rho$ is the ion mass density.
We take a standard chemical composition ($X=0.75, Y=0.25$)
in the corona and the average molecular weight then $\mu_0=0.62$.
In a fully ionized gas of standard composition the electron number
density is about 10\% larger than the ion number density. The last
expression for the total pressure in Eq. (1) is chosen for
simplicity. For $\mu_0=0.62$ it is exact if $T_{e}=T_{i}$, but it
overestimates $P$ for $T_{e} \ll T_{i}$ by 5\%. We neglect this difference.

From the conservation of mass we get our first equation
\begin{equation}
\frac{d}{dz}(\rho v_z)=\eta_M \cdot \frac{2}{r}\rho
 v_r -\frac{2z}{r^2+z^2}\rho v_z,
\end{equation}
with $v_r$ the radial diffusive velocity and $\eta_M$ the mass
advection modification term (see next section). The second term
in Eq. (2) approximately takes into account the effect of the changing
channel cross section $\approx (1+z^2/r^2)$, for the ascending flow as
its shape changes from cylindrical to spherical at large height $z$
(compare Meyer et al. 2000).

The second equation is the z-component of the equation of motion,
\begin{equation}
\rho v_z \frac{dv_z}{dz}=-\frac{dP}{dz}-\rho \frac{GMz}{(r^2+z^2)^{3/2}}.
\end{equation}
with $G$ gravitational constant and $M$ mass of the central black
hole. The r-component of the equation of motion to first order gives
the rotational velocity $v_\varphi$ as equal to the local Kepler
velocity, $v_\varphi^2=GMr^2/(r^2+z^2)^{3/2}$, neglecting radial
pressure gradients, small of order $\Re Tr/GM\mu \leq
\frac{1}{5}$ in our
solutions. $T$ stands for the temperature which determines the
hydrostatic layering, $T\approx T_i$. The $\varphi$-component
yields angular momentum conservation and determines $v_r$.

The vertical conductive heat flux provides our third equation
\begin{equation}
F_c=-\kappa_0T_e^{5/2}\frac{dT_e}{dz},
\end{equation}
with $\kappa_0=10^{-6}$g cm  ${\rm s}^{-3} {\rm K}^{-7/2}$ for a fully
ionized plasma (Spitzer 1962). This formula for the heat conduction is
valid if the electron mean free path is small compared to the length
over which the temperature changes significantly. In the region of our
solution where heat conduction is a significant contribution this is
the case (occasionally at least marginally).

The two remaining equations, the energy equations for ions and for
electrons, as derived by Liu et al. (2002), 
include the cooling and heating processes in the hot
corona and the equations are now modified for the multi-zone model.
For ions the energy balance is determined by viscous
heating, cooling by collision with electrons and radial and vertical
advection. The friction is taken proportional to the pressure with a
standard $\alpha$-prescription.

\begin{eqnarray}
\frac{d}{dz}(\rho_{i} v_zu_i) & = 
&\frac{3}{2}\alpha P\Omega- q_{ie}
+\eta_{\rm E}\cdot \frac{2}{r}\rho_{i} v_r 
u_i  \nonumber  \\
& & -\frac{2z}{r^2+z^2}(\rho_{i}
v_z u_i),
\end{eqnarray}
with $\Omega$ the rotational frequency, $\rho_i$ ion density,
$\alpha$ viscosity
parameter.
$u_i$ is the specific energy of ions
\begin{equation}
u_i=\big(\frac{v^2}{2}+\frac{\gamma}{\gamma-1}\frac{P_{i}}{\rho_{i}}
+ \Phi\big),
\end{equation}
with $\Phi$ the (Newtonian) potential 
\begin{equation}
\Phi=-\frac{GM}{(r^2+z^2)^{1/2}}.
\end{equation}
$\frac{3}{2}\alpha P\Omega$ is the viscous heating rate per unit volume  
and ${q_{ie}}$ the rate of energy transfer from ions to electrons
(Stepney 1983) as taken in Liu et al. (2002). The energy advection
modification term $\eta_E$ is discussed in the next section.
The contribution of the frictional stress to the divergence of the
energy flow is already explicitly taken into account as frictional
dissipation term in the equations.

For the multi-zone model there are two changes compared to
the one-zone model, (1) the mass and energy modification factors
and (2) we now take into account the radial dependence of
$v_\varphi^2$ in the formula for the specific energy (see Sect. 2.3.3).

For electrons the energy balance is determined by the processes of
heating by collisions with ions, cooling by bremsstrahlung
(free-bound and free-free transitions), Compton cooling,
and vertical thermal conduction.We neglect here the radial
thermal conduction in a first approximation. In the main part of the
vertical structure solution it is not a dominant term
(see Meyer et al. 2000).

The second energy equation is
\begin{eqnarray}
\frac{d}{dz}(\rho_{e} v_z u_e +F_{c}) &
= &q_{ie}-n_e n_iL(T_e)-{q_{\rm {comp}}}
  \nonumber \\ & &
+\eta_{\rm E}\cdot\frac{2}{r}\rho_{e} v_r u_e 
  \nonumber \\ & &
 -\frac{2z}{r^2+z^2}(\rho_{e}
v_z  u_e +F_{c} ),
\end{eqnarray}
where $u_e$ is specific energy of electrons (defined correspondingly to
$u_i$) and $n_e n_i L(T_e)$ is 
the bremsstrahlung cooling rate and ${q_{\rm {comp}}}$  the Compton
cooling rate (Rybicki \& Lightman 1979), as in Liu et al. (2002).

In the energy balance for the electrons one could omit the terms
multiplied with $\rho_e$ due to the low ratio of electron to ion mass.
We keep the equation for ions, Eq.(5), and replace the equation for
electrons by a joint equation for electrons and ions
(with $u$ the specific energy of electrons and ions).

\subsection{The modification terms for radial inflow/outflow}

We first derive the formula for the radial diffusive velocity and then 
the mass and energy advection modification terms. For simplicity these
modifications are taken in a height averaged way.

\subsubsection{The radial diffusive velocity}
The equation for local angular momentum conservation is
\begin{eqnarray}
\frac{\partial}{\partial r}(2\pi r\rho v_r r^2\Omega) +\frac{\partial}{\partial
z}(2\pi r\rho v_z r^2\Omega)
+\frac{\partial}{\partial r}(2\pi r\tau_{r\varphi})=0
\end{eqnarray}
with the frictional stress $\tau_{r\varphi}$ 
\begin{equation}
\tau_{r\varphi}=-\alpha P= \mu r\frac{\partial \Omega}{\partial r}
\approx-\frac{3}{2}\Omega \mu,
\end{equation}
$\mu$ viscosity. For the lower, dominant part of the corona it is a
reasonable approximation to neglect the $z$-dependence of $\Omega$ (for an
error estimate see Meyer et al. 2000). 
Mass conservation 
\begin{equation}
\frac{\partial}{\partial r}(2\pi r\rho v_r)+\frac{\partial}{\partial z}
(2\pi r\rho v_z)=0
\end{equation}
together with $\frac{\partial}{\partial z}(r^2\Omega)
\approx 0$ yields
\begin{equation}
2\pi r\rho v_r \frac{\partial}{\partial z}(r^2\Omega)+
3\pi \frac{\partial}{\partial r}(r^2\Omega \mu)=0.
\end{equation}
Using 
\begin{equation}
\frac{\partial(r^2\Omega)} {\partial r}\approx
\frac{1}{2}\frac{r^2\Omega}{r}
\end{equation}
one finally obtains for the radial diffusive velocity (inward directed)
\begin{equation}
v_r  = 
-\frac{3\alpha P}{r\Omega\rho}(1+2\frac{\partial \ln \mu}
{\partial \ln r}) 
= - \frac{\alpha V_s^2}{v_{\rm K}} (1+
2\frac{\partial \ln \mu}
{\partial \ln r}),
\end{equation}
with $v_{\rm K}$ Kepler velocity. As a simplification
we take the rate of change of
the viscosity the same at all heights $z$ and replace 
the logarithmic derivative of $\mu$ by the
logarithmic derivative of its integral $f$
\begin{eqnarray}
\frac{\partial\ln\mu}{\partial \ln r}=\frac{d\ln f}{d\ln r},
\hspace{1cm} f=2 \int_{z_0}^{z_1}\mu dz
\end{eqnarray}
(integration over the height of the corona on both sides of the disk).
$f$ results from the computed vertical distribution of the
pressure according to Eq. (10). 
With this replacement Eqs. (10), (14) and (15) yield for the coronal
accretion rate (positive if
inward)
\begin{eqnarray}
\dot M_c &=& 2 \int_{z_0}^{z_1} 2\pi r \rho v_r dz=
3\pi f(1+2\frac{d\ln f}{d\ln r})
\nonumber \\
& = & -2\pi r\Sigma \overline{v}_r
\end{eqnarray}  
the third equality defining the mean radial velocity $\overline{v}_r$,
with the surface density $\Sigma=2 \int_{z_0}^{z_1}
\rho dz$ (both sides of disk). 

\subsubsection{The divergence of mass and energy flow}
We derive the modifications of the
sidewise advection terms in the equations for conservation of
mass and energy again as mean values over $z$.
The radial divergence is written as
\begin{equation}
 \frac{1}{r}\frac{d}{dr}\big(r\Sigma \overline{v}_r \big)
  = -\frac{2}{r} \Sigma \overline{v}_r \eta_M
\end{equation}
such that $\eta_M$ measures the radial divergence in terms
of its one-zone model value.  Thus follows
\begin{equation}
 \eta_M=-\frac{1}{2}\frac{|d\ln(r \Sigma \overline{v}_r|)
}{d\ln r}=-\frac{1}{2}
\frac{d\ln |\dot M_c|}{d\ln r}.
\end{equation}
With Eq. (16)
the mass modification factor can also be written in the form
\begin{equation}
\eta_M=-\frac{1}{2} \frac{d\ln |f(1+2\frac{d\ln f}{d\ln r})|}{d\ln r}.
\end{equation}
The divergence of the energy advection involves the derivative of
$\dot M_c \cdot u(r,\eta)$ with $u$ the  specific energy (see below), 
\begin{equation}
\eta_E=-\frac{1}{2} \frac{d\ln |\dot M_c(r,\eta)\cdot u(r,\eta)|}{d\ln r}.
\end{equation}

The potential, kinetic, and thermal specific energies all scale
approximately as $1/r$ (see Liu et al. 2002, Fig. 4). The factor then is
\begin{equation}
\eta_E= \eta_M +0.5.
\end{equation}

\subsubsection{The specific energy}
In the specific energy $u$ the kinetic energy
is $\frac{v^2}{2}=(v_r^2+v_z^2+v_\varphi^2)/2$. Since $v_\varphi$,
the dominant term, is $r$ dependent, and it has to be taken into account as
$v_\varphi^2=GM r^2/(r^2+z^2)^{3/2}$ (In the one-zone model it was
taken as constant and dropped out of the equations).

\subsection{Boundary conditions}
The boundary conditions are the same as those in the one-zone model.

As the lower boundary of our calculations at $z=z_0$, we take the 
level where electron temperature and ion temperature are already about
the same and have the value $T_e=T_i=10^{6.5}$K. Below an analytic
solution suffices to cover the small remaining extent before the disk
chromosphere is reaching it and yields a relation between pressure and
downward heat flux (for a more detailed discussion see Meyer et al. 2000). 

At the upper boundary with no pressure confinement at
infinity we require sound transition at some height
$z=z_1$ (free boundary), $v_z=V_{\rm s}$. Further with no influx of
heat from infinity there, we require $F_c=0$ neglecting a small
remaining outward heat flow..

This constitutes 5 boundary conditions for the 5 ordinary differential
equations in $z$.

\section{A consistent solution at all distances $r$}
The equations described are valid for accretion in different
astrophysical objects, stellar and galactic black hole masses.The
solutions scale with Eddington accretion rate and Schwarzschild
radius. In the following we present the solutions for an accretion flow
around a $6 M_\odot$ black hole (the Eddington accretion rate then is
$8.4\cdot 10^{18}$g/s, the Schwarzschild radius $1.8\cdot10^6$cm).
We take the viscosity parameter $\alpha$=0.3 (for the
dependence of the coronal structure on the viscosity see
Meyer-Hofmeister \& Meyer 2001).

\subsection{How to find consistent solutions}
The main goal of the new approach is to find out how the
corona is stretched out above the disk in radial extent.
The one-zone model gives a
description of the coronal structure only for the one inner zone
where the thin disk underneath becomes truncated.
This is a reasonable first approach when the evaporation efficiency
is steeply increasing with decreasing $r$ and evaporation is the
dominant mass feeding cause there. (Mass flow into the corona from
farther outward was neglected and 
simple estimates for the flow of mass, energy and angular momentum
through this one zone were made). 

For the determination of the coronal mass flow in a radially extended
region the difference of inflow and
outflow in a succession of radial zones is now properly accounted for. 
This means that the coronal mass flow at a certain radius depends on the
coronal structure in the neighboring zones and must be
determined iteratively.

We first compute the coronal structure for
successive distances and a series of values for the modification terms
to construct a grid of curves $f(r,\eta_M,\eta_E)$.
The derivative of $f$ only appears in $v_r$
and in the modification terms. We therefore take these terms together
as
\begin{eqnarray}
\tilde\eta_M=\eta_M \cdot(1+2\cdot \frac{d\ln f(r,\eta)}{d\ln r})
\nonumber \\
\tilde\eta_E=\eta_E \cdot(1+2\cdot \frac{d\ln f(r,\eta)}{d\ln r})
\end{eqnarray}

We show some of these results in Fig. 1. A value $\tilde{\eta}_M=0$ means
a net zero outflow. In this case we found that the structure 
approximately is independent of the value of $\tilde{\eta}_E$.
The value of the viscosity
integral  $f$ decreases with increasing net flow. Higher values of $f$
correspond to higher values of the pressure at the bottom of the
corona. The value $\tilde\eta_M=1$ corresponds to the net outflow taken in the
one-zone model (aside from the now included term $v_\varphi^2$).
From Fig. 1 we see that advection of energy has a different effect
on the coronal structure at different distances $r$. This depends on whether
the frictional energy release preferentially goes into radiation or
into advective loss (compare Fig. 5 Meyer et al. 2000).

\begin{center}
\begin{figure}[ht]
\includegraphics[width=7.5cm]{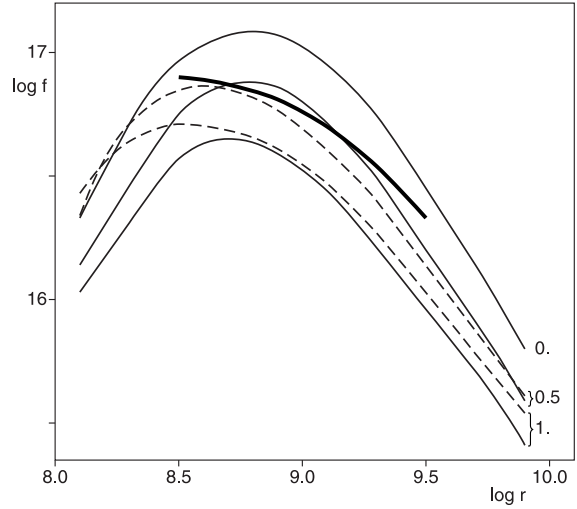}
\caption {Values of the viscosity integral
$f(r,\tilde\eta_M,\tilde\eta_E)$ for different parameters, basic curves for
the evaluation of the coronal flow. solid lines: $(\tilde\eta_M,
\tilde\eta_E)$=(0,0),
(0.5,0.5),(1,1); long dashes: $(\tilde\eta_M, \tilde\eta_E)$=(0.5,1),(1,1.5)
For (0,0.2) the line is approximately the same as for
(0,0). Thick line: example of a consistent curve log $ f$, including the mass,
energy and angular momentum flow consistently (in our computations we
investigate the accretion flow around a $6 M_\odot$ black hole).
}
\end{figure}
\end{center}

What we finally need is a series of coronal structures
at each distance determined for given $\tilde\eta$ values
($\tilde\eta$ indicates the dependence on both
$\tilde\eta_M$ and $\tilde\eta_E$) so that the curve
log $ f(r,\tilde\eta)$ has the slope
$s=d\ln f/d\ln r$ that fits to these values taken for
$\tilde\eta$. From Eqs.(19), (21), and (22) and the definition of
$s$ we obtain

\begin{equation}
\tilde\eta_M  = -\frac{1}{2}
(s+s^2+2\frac{ds}{d\ln r}),\hspace{0.3cm}
\tilde\eta_E  =(\tilde\eta_M+0.5)(1+2s).
\end{equation}

This implies an ordinary differential equation of second order for $f(r)$.
A unique solution then requires two boundary conditions, e.g. at
an inner and an outer edge. The solutions thus form a two-dimensional
manifold. Different approaches might be possible.
An iterative procedure without enforcing appropriate boundary
conditions, starting 
from one of our curves in Fig. 1, determining the derivatives of
log $f$ and the $\tilde\eta$ values and then the appropriate new
log $f$ curves, seems not to converge to a consistent solution. 
This is probably due to the large number of possible
solutions corresponding to the diversity of possible boundary conditions.

We use the following procedure.
We start at a chosen distance $r=r_{\rm{tr}}$ where the
thin disk is truncated. There the boundary condition is
$s(r_{\rm {tr}})=0$ which means that all angular
momentum carried inward into the inner disk free region is returned by friction
(see discussion in Meyer et al. 2000). Now we construct a consistent
log $ f$ curve step after step. Step 1: we choose an initial value of 
log $ f(r,\tilde\eta)$ (from our computed grid of curves);
from Eq. (23) we see which value has to be chosen for $\frac{ds}{d\ln r}$
to give the values $\tilde\eta_M$ and $\tilde\eta_E$ that belong to
the chosen initial value of $f$
(the derivative is approximated by a difference quotient
$(s(log r_{\rm {tr}}+\Delta log r)$ - $s(log r_{\rm {tr}})/\Delta log
r$, we used
$\Delta log r$=0.2). This determines the slope $s$ with which to
proceed to the next log $ f$ value at the next radius. Step 2:
the slope determined in step 1 yields the next value for log $ f$;
again we determine the further slope so that relation (23) is
fulfilled.  Each step thus determines the further outward slope $s$
of the log $ f$ curve so that the
$\tilde\eta$ values and $f(r,\tilde\eta)$ are consistent.

For a given truncation radius and for each initial value of log $f$
we get with this procedure one consistent log $ f(r,\tilde\eta)$ 
curve. A second
boundary condition at an outer boundary of the interval for which we
determine the coronal flow allows to discriminate between
the various curves that belong to one truncation radius. The slope at 
the outer distance is a possible boundary condition. For
$s= 1/2$ the radial diffusive velocity $v_r$ becomes zero, no mass flow
inward or outward. Then closer to the black hole the mass flows
inward, at larger distance the mass flows outward. In different disks
the outer boundary condition might be different. In the case
of X-ray binaries it seems plausible that tidal forces
at the disk boundary prevent outer coronal mass outflow, that is
$d\ln f/d\ln r$ becomes -1/2 there.

\subsection{Computational results}
Our method yields the coronal structure solutions for
given truncation radii. This gives the mass flow rate in the corona at the
truncation radius where no mass flows in the cool disk anymore.
This mass flow together with the wind loss from the corona added up
over the radial extent of the corona is equal to the mass flow in
the far out cool disk. This establishes the dependence of the disk
truncation radius on the mass flow rate in the outer thin disk.
In Fig. 1  we show one log $ f$ curve starting at
log $r_{\rm{tr}}$=8.5 In Fig. 2 we show three characteristic curves
starting at log $r_{\rm{tr}}$=8.5 as well as the characteristic
run of log $f$ curves for other
truncation radii. Generally the lower the initial of log $ f$ the
steeper is the corresponding log $ f$ curve.
One also sees that a small variation in the initial value of $ f$ 
yields a large variation of the radius where $v_r=0$ (marked by a
diamond in Fig. 2). This indicates a somewhat
weak dependence of the inner coronal mass flow
rate on the outer boundary condition.

\begin{center}
\begin{figure}[ht]
\includegraphics[width=8.cm]{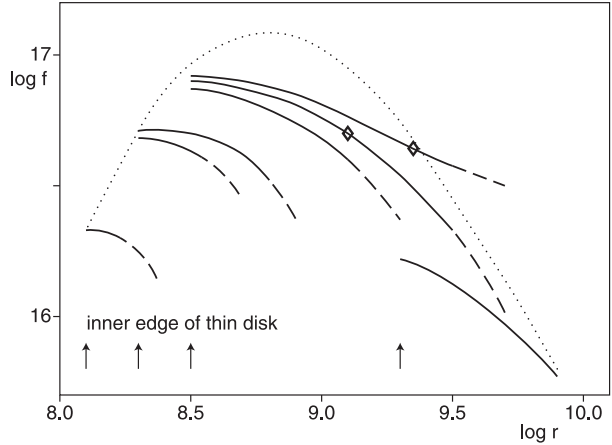}
\caption {Consistent solutions for log $ f$ for truncation radii
log $ r$=8.1 to 9.3. Diamonds on two curves mark where $v_r=0$.
The coronal flow at the truncation radius is proportional to $f$. 
Dashed parts indicate uncertain extrapolation
outside of the computed basic grid curves. The dotted
curve is identical with the uppermost curve in Fig. 1 (accretion flow
around a $6 M_\odot$ black hole)
}
\end{figure}
\end{center}

Comparing the $f$ curves of the solutions with the field which our 
basic grid curves cover in the log $ f$-log $ r$ diagram we see that the
solution $f$- curves can easily lead into areas outside of the grid,
either to  higher or lower values of log $ f$. If they stay within the
range of computed basic curves necessarily the slope becomes less than
-1/2. No consistent solutions have been constructed which have $v_r$=0
at a very large distance. The too steeply decreasing curves 
correspond to the situation that the coronal flow is inward only near the
inner boundary but turns into an outflow already a short distance
away from it. To continue these
curves correctly we need the coronal structures for high $\tilde\eta_M$
values, which means a high net radial mass outflow. 
The flatter curves also can leave the field of standard solutions, that is
$\tilde\eta_M\leq 0$. The corona would then have a negative radial 
net outflow, which means less mass radially leaves the zone at
the inner radius than enters into it at the outer radius.

\begin{figure}[ht]
\includegraphics[width=8.8cm]{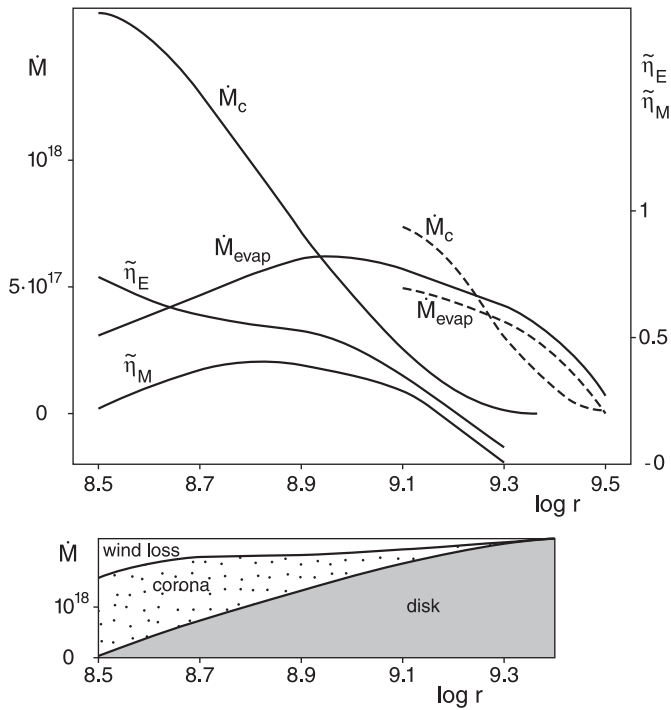}
\caption {Upper panel: solid lines: $\dot M_c$ coronal mass flow rate
above the thin disk around a $6 M_\odot$ black hole,
truncated at log $ r_{\rm{tr}}$=8.5, mass and energy flow modification values
$\tilde\eta_M$ and $\tilde\eta_E$. $\dot M_{\rm{evap}}$ evaporation rate
(see text). Dashed lines: $\dot M_c$ and $\dot M_{\rm{evap}}$ for
disk truncation at log $ r_{\rm{tr}}$=9.1. 
Lower panel: mass flow rate in the thin disk (gray area) and in the
corona (dotted area) as function of distance $r$ from the black hole,
area remaining above indicates the rate of gas loss in the wind from
the corona.
}
\end{figure} 
In Fig. 3 we show the coronal mass flow for the case of 
a thin disk with a mass flow rate of $2.4\cdot 10^{18}$g/s from outside.
This disk becomes truncated at log $ r_{\rm{tr}}$=8.5.
We show how the coronal flow
increases inward (together with the mass and energy flow modification values
$\tilde\eta_M$ and $\tilde\eta_E$ of this consistent solution).
The evaporation
rate measured by $\dot M_{\rm{evap}}=2\pi r^2 \dot m_0$ ($\dot m_0$
vertical mass flow rate per unit surface at the bottom of the corona,
in units of $\rm{g/cm^2s}$) gives the amount of gas locally evaporated
into the corona. The earlier one-zone
model only considered evaporation in the innermost region. Here
our solution shows a radially quite extended evaporation region.
In the lower panel of Fig. 3 we show how the mass flow in the thin
disk decreases inward. Gas evaporates into the hot coronal flow, from
which a significant part, $\approx 50\%$ is lost in a wind. 
 
In Fig. 4 we show the results from a series of computations with 
successive truncation radii.
We chose solutions where $v_r$=0 occurs relatively far outward and give
the coronal mass flow rates at the truncation radii. In addition we
show the virial and ion and electron temperatures of the corona at
height $z=r$ at the truncation radius.
Note the small difference between $T_i$ and $T_e$. The coronal mass flow
rates are higher than the ones derived with the one-zone model. This
is in part due to inclusion of the value of $v_{\varphi}^2$ in the
specific energy.

\begin{center}
\begin{figure}[ht]
\includegraphics[width=7.cm]{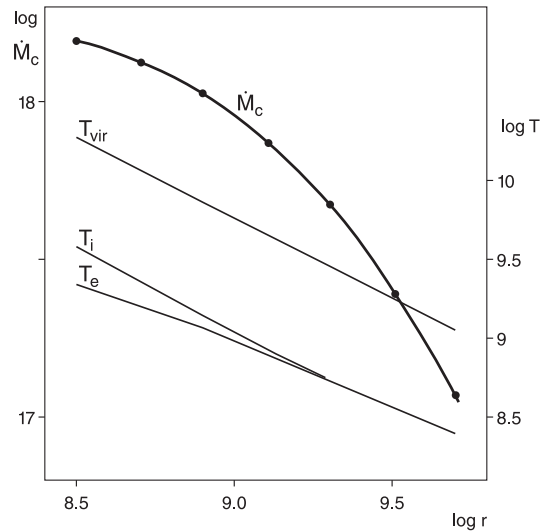}
\caption {Thick line: series of solutions for the coronal mass flow
rate $\dot M_c$, for disk truncations at log $ r$ from 8.5 to 9.7;
thin lines: ion and electron temperature at coronal height $z=r$ together
with the virial temperature $T=\frac{\mu}{\Re}\frac{GM}{r}$.}
\end{figure}
\end{center}

\section{Discussion and conclusions}
\subsection{The nature of the solutions - wind loss}
Having developed the procedure for solving the differential equations
described above we see that there are quite different regimes
of solutions. Up to now we have studied solutions with a moderate advective
mass and energy net outflow. These solutions seem adequate to describe
the coronal flow above a thin accretion disk e.g. in X-ray binaries. For
the investigation of AGN disks another outer boundary condition might
be taken, but this would not affect the solutions essentially.
It will be interesting also to study other regimes of solutions i.e. 
solutions leading into the area above our network of curves
log $ f(r,\tilde\eta_M,\tilde\eta_E)$ in the log $f$-log $r$ diagram.
Test computations showed that in these solutions an increasing part
of the coronal flow does not proceed inward but is lost in a wind.
In our equations flaring vertical columns are considered. If the wind
loss becomes important the structure becomes dependent on the assumed
geometry. This geometry then has to be considered carefully to have a
consistent picture how the expanding wind structure is arranged
radially. This is an interesting topic in view of the question how
much gas actually enters the disk free inner region and flows towards
the black hole.

\subsection{Disk truncation closer to the black hole?}
If we imagine that the mass flow rate in the thin disk
can be arbitrarily high we should encounter a situation where the
disk is not yet truncated at e.g. log $ r$=8.5. If evaporation
continues a truncation farther inward is possible. We have not yet
found a consistent solution for the coronal flow corresponding to a disk
truncation farther in. As can be seen from our solutions starting at
larger distances we would have to start with higher initial values of
log $ f$, values for which we have no log $ f$ curves. Logically, a coronal
structure for condensation of gas instead of evaporation should then be
appropriate. Such condensation already appears in the algebraic
treatment of coronal structure by R\'o\.za\'nska and Czerny (2000a).
Thus in our approach which consistently derives advective terms from
the radial gradients condensation should also be expected. In the
case of a net mass inflow by sideways advection part of the gas flows
down into the cool disk while another part flows upward and leaves as
a wind.

\subsection{Compton effect}
The computed coronal structure is valid for a mass flow rate in the
thin disk low enough not to affect the corona by Compton cooling.
Results in the work of Liu et al. (2001, Fig. 5) have shown that 
for higher mass flow rates the Compton cooling leads
to a steep decrease of the evaporation rate inwards. For our investigation
this means that then the basic grid of log $f$ curves (Fig. 1)
would also have
a more pronounced decrease towards smaller $r$. This is the region
where we have not yet constructed consistent solutions. Having
solved the problem of solutions for smaller $r$ it is meaningful
to discuss the effect of Compton cooling on consistent coronal flow
solutions.

\subsection{Spectral transitions in X-ray binaries}
The two examples for coronal flow corresponding to disk truncation at 
log $ r_{\rm{tr}}$ equal to 8.5 and 9.1 (see Fig. 3) show the
difference in the mass flows in disk and corona.
For a mass flow of about $9\cdot 10^{17}$g/s 
the thin disk becomes truncated at log $ r_{\rm{tr}}$ =9.1. 
The hard spectrum is determined by this amount of mass flow 
in the corona, farther inward in the ADAF. If the mass flow in the
thin disk is 
$2.4\cdot 10^{18}$g/s the disk is truncated at
log $ r_{\rm{tr}}$= 8.5 (compare Fig. 3 for the amount of wind loss),
the spectrum correspondingly changed.
A factor of about 2.5 in the mass
flow rate might occur during the increase or decrease to/from an
outburst in an X-ray transient (even larger if wind loss is included). 
This leads to the change of the
spectrum. For very high rates the disk is not truncated. For this case
we can not yet present a solution.

\subsection{Concluding remarks}
The newly developed method for the construction of radially
extended coronal flows allows a better description for the change
from accretion via a thin disk to a disk free pure coronal flow
(ADAF). We have presented results for disk truncation at various
distances $r$ from a stellar black hole of $6 M_\odot$ applicable
in X-ray binaries. The method also allows
to study other regimes of solutions and might shed light on wind
loss under those conditions.
Particularly important will be to investigate whether the corona 
can partially condense into the thin disk again or keeps flowing at a
significant rate. The solutions scale with
Eddington accretion rate and Schwarzschild radius, are therefore also 
relevant for the accretion flow in AGN.

\end{document}